\documentclass[manuscript]{aastex631}

\usepackage{ulem}
\usepackage{float}
\usepackage{graphicx}
\usepackage{amsmath}
\usepackage{blindtext}
\usepackage[hyphens]{url}
\usepackage{multirow}
\usepackage{longtable}
\usepackage{hyperref}
\hypersetup{
colorlinks = True,
linkcolor = blue,
citecolor = blue
}

\begin{document}

\title{Deconstructing the Properties of Solar Super Active Region 13664 in the Context of the Historic Geomagnetic Storm of 2024 May 10-11}

\author[0009-0009-4297-1861]{Priyansh Jaswal}
\affiliation{Center of Excellence in Space Sciences India, Indian Institute of Science Education and Research Kolkata, Mohanpur, 741246, India}

\author[0000-0001-7229-5192]{Suvadip Sinha}
\affiliation{Center of Excellence in Space Sciences India, Indian Institute of Science Education and Research Kolkata, Mohanpur, 741246, India}

\author[0000-0001-5205-2302]{Dibyendu Nandy}
\affiliation{Center of Excellence in Space Sciences India, Indian Institute of Science Education and Research Kolkata, Mohanpur, 741246, India}
\affiliation{Department of Physical Sciences, Indian Institute of Science Education and Research Kolkata, Mohanpur, 741246, India}

\correspondingauthor{Dibyendu Nandy}
\email{dnandi@iiserkol.ac.in}

\begin{abstract}
Active regions (ARs) are sites of strong magnetic fields on the solar surface whose size can be several times that of Earth. They spawn dynamic activity sometimes resulting in severe space weather. Some ARs characterized by extraordinary magnetic properties, and exhibiting extreme activity, are termed as super active regions (SARs). Recently, solar AR 13664 produced 23 X-class flares and unleashed multiple coronal mass ejections which triggered a severe geomagnetic storm during 2024 May 10-11 -- the strongest storm on record since 2003. Here we put AR 13664 in historical context over the cumulative period of 1874 May-2024 June. We find that AR 13664 stands at 99.95 percentile in the distribution of area over 1874 May-2024 June, and at 99.10 percentile in terms of flux content among all ARs over the period 1996 April-2024 June. Our analysis indicates that five of its magnetic properties rank at 100.00 percentile among all ARs observed during 2010 May-2024 June by the Solar Dynamic Observatory. A total of 16 magnetic properties of AR 13664 are ranked higher than 99.00 percentile when compared to other ARs recorded in SHARP data series which includes all well established flare relevant parameters. Furthermore, we demonstrate that AR 13664 reached its most dynamic flare productive state following a rapid rate of rise of its flare-relevant parameters and that the X-class flares were more frequent near their peak values. Our analyses establish AR 13644 to be a super active region and provide a paradigm for investigating their flare-relevant physical characteristics.
\end{abstract}

\section{Introduction}
The Sun is an active star. Its dynamics and interaction with planetary environments are governed by the solar magnetic activity \citep{Usoskin_2023_LivRevSolPhys}. The genesis of the solar magnetic fields owes to the dynamo mechanism deep within the interior of the Sun \citep{Nandy_2002_Science, Hazra_2016_ApJ, Charbonneau_2020_LivRevSolPhys, Hazra_2023_SpaceSciRev}. These magnetic fields buoyantly rise up and protrude through the photosphere into the solar atmosphere, resulting in the formation of regions of intense magnetic fields called active regions (ARs). On the solar surface, when observed in white light, these ARs are accompanied by dark patches called sunspots \citep{VanDriel-Gesztelyi_2015_LivRevSolPhys}. The magnetic field evolution and instabilities of AR flux systems, their interaction and reconnection with nearby magnetic fields are believed to drive phenomena such as flares and coronal mass ejections (CMEs) \citep{Cook_2009_ApJ, Yeates_2010_ApJ, Zhu_2020_ApJ, Georgoulis_2024_AdvSpaceResearch}. When Earth-directed, flares and CMEs can lead to extreme near Earth space weather and geomagnetic storms \citep{Roy_2023_ApJL} posing a hazard to our space based technological assets \citep{Schrijver_2015_AdvSpaceResearch, Nandy_2023_JASTP, Baruah_2024_SpaceWeather}. However, around 40\% of the strong flare events occur from only (approximately) 0.5\% of the ARs throughout a solar cycle \citep{Chen_2012_aanda}. One of the foremost characteristics of such highly active ARs is huge area, and they are referred to as super active regions (SARs) \citep{Jiang_2016_ApJ}. It is known that larger ARs preferentially emerge during the maxima phase of a sunspot cycle \citep{Chen_2012_aanda}; with Sun ramping up towards the maxima of ongoing sunspot cycle which is predicted to be stronger than sunspot cycle 24 and to occur in 2024, more SARs are expected to emerge \citep{Bhowmik_2018_NatComm, Jaswal_2023_MNRASL, Jha_2024_ApJL, Pal_Shao_2024_MNRAS} raising expectations for cascading space weather events \citep{Oliveira_2024_SpaceWeather}. Therefore, in order to enhance our forecasting capabilities and better understand space weather it is important to study and keep a track of such SARs and identify why they are unique. 

\cite{Tian_2002_SolPhys} studied twenty five most violent SARs during solar cycles 22 and 23, and showed that most of the SARs tend to have larger net magnetic flux and abnormal magnetic topology. Further they suggested that there are certain longitude bands on the solar surface where complex ARs emerge which can easily host major solar storms. In a study of SARs during the descent phase of sunspot cycle 23, \cite{Romano_2007_aanda} found that magnetic polarity imbalance and length of magnetic inversion line are linked to the triggering of solar flares. Findings by \cite{Dhakal_2023_ApJ} tell that complex magnetic structure, longer strong gradient polarity inversion line and greater total unsigned flux near the polarity inversion line (R\_VALUE) are key indicators of high flaring activity in ARs. Another recent study by \cite{Brooks_2021_ScienceAdv} showed how SAR 11944 led to a major solar energetic particle event during the maxima phase of sunspot cycle 24.

\begin{figure*}
    \centering
    \includegraphics[width=0.95\linewidth]{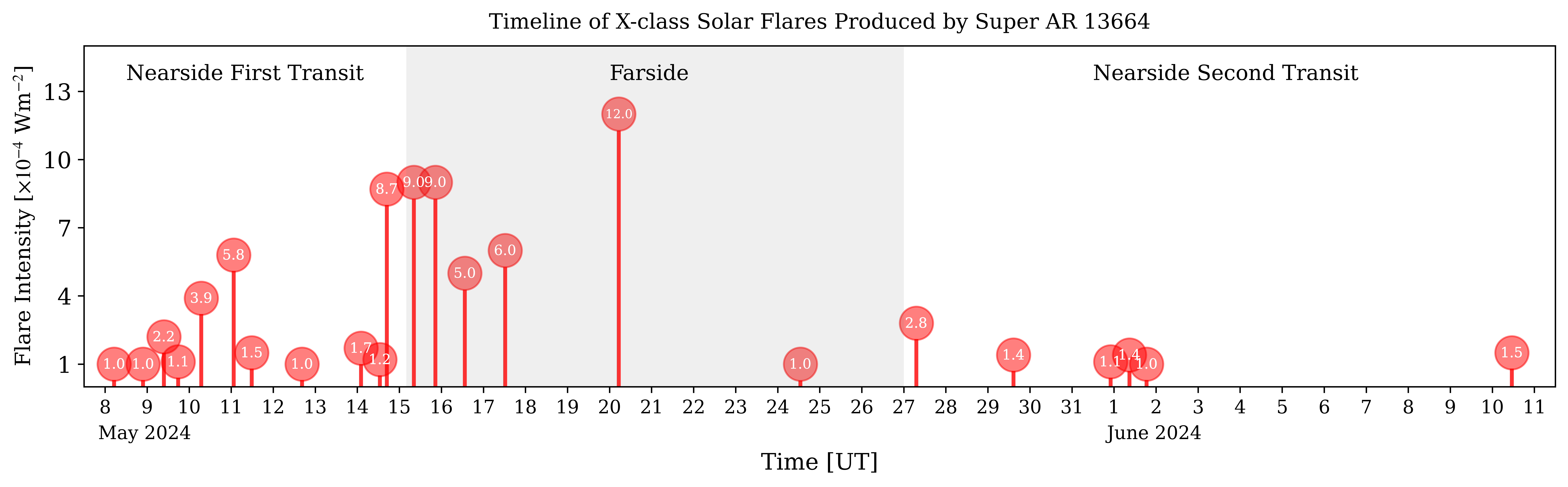}
    \caption{Timeline of all the X-class solar flares produced by AR 13664 during its first nearside transit, farside transit, and second nearside transit on the solar disk. During its farside transit it happened to be in the field of view of the instruments onboard the Solar Orbiter (SolO). The nearside and farside observations -- delineated by the shaded region -- are marked in red circles wherein the inscribed numbers represent the intensity of the respective X-class flare produced by the super active region. \textit{Source}: \href{https://www.esa.int/Science_Exploration/Space_Science/Solar_Orbiter/Can_t_stop_won_t_stop_Solar_Orbiter_shows_the_Sun_raging_on}{ESA/SolO}}
    \label{fig:Figure 1}
\end{figure*}

A very recent active region AR 13664 produced 23 X-class solar flares and multiple CMEs in the course of its first nearside transit, farside transit, and second nearside transit on the solar disk over the period 2024 May 01-June 12 (refer to Figure \ref{fig:Figure 1}). Some of these CMEs which emanated from AR 13664 hit Earth during 2024 May 10-11, creating the most intense geomagnetic storm of the past two decades whose impacts are just beginning to be appreciated \citep{Witze_2024_nature, Hajra_2024_ApJ}. The aim of this study is to establish how rare of an active region AR 13664 is in the historical context and explore what properties could have led to its super activity.

There are several dimensions to this work. We ascertain whether AR 13664 qualifies as a super active region in terms of historical observations, explore how the flare-relevant properties of AR 13664 compare to those observed earlier and what is the evolution of few highly flare-relevant magnetic parameters during the visible (Earth facing) lifespan of this active region.  The first task is further subdivided to 1) area observations going far back in history to 1874 because other parameters were not available, 2) flux comparison going back to 1996 (MDI-HMI era), and 3) comparison of flare relevant parameters \citep[including non-potentiality markers,][]{Hahn_2005_ApJ} available systematically from vector magnetic fields in the HMI era.

\section{Methods and Results }

\begin{figure*}
    \centering
    \includegraphics[width=1\linewidth]{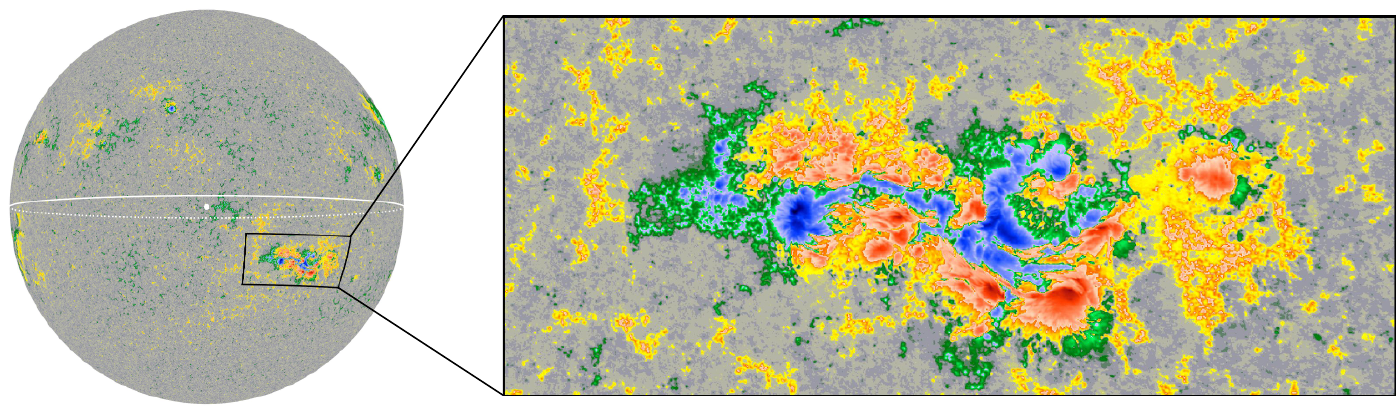}
    \caption{[Left] Line of sight (LOS) full disk magnetogram of Sun on 2024 May 09 at 17:36:00 UT gleaned from SDO/HMI. White solid (dotted) curve depicts the equator of Sun on the nearside (farside), and white spot represents the center of the Sun. [Right] Cylindrical equal area (CEA) projected LOS magnetogram for Space-weather HMI Active Region Patch -- numbered 11149 (with AR 13664 as the primary active region) -- at the instance when the area of active pixels (AREA\_ACR) was maximum (4394.67 \(\mu\)Hem) during its first nearside transit on the solar disk over the period 2024 Apr 30-May 15. The complex configuration of this AR consisting of elongated positive (blue) polarity nested in between background negative (red) polarity belongs to the \(\beta\)-\(\gamma\)-\(\delta\) Hale class.}
    \label{fig:Figure 2}
\end{figure*}

As evident from Figure \ref{fig:Figure 2} the magnetic configuration of the active region is highly complex with elongated positive polarity patches nesting within a large scale distribution of negative polarity patches, resembling the \(\beta\)-\(\gamma\)-\(\delta\) Hale class; this allows for strong opposite polarities to coexist in close proximity -- favoring magnetic reconnection in the overlying flux system. The evolution of the regions surrounding AR 13664 suggest a nearby flux emergence which was officially assigned an identification number by the National Oceanic and Atmospheric Administration (NOAA) -- AR 13668 -- on 2024 May 06. However, within a day the flux system of AR 13668 coalesced with that of AR 13664's and all the energetic X-class flare activity commenced after 2024 May 07 by when merging of these two NOAA ARs had already occurred. They appear within the same Helioseismic Magnetic Imager (HMI) AR patch (HARP), numbered 11149; therefore for brevity, henceforth we denote this AR complex bounded by HARP number 11149 as AR 13664. To get an idea of how large AR 13664 is as compared to all the sunspots recorded in the past, we use sunspot area data maintained by the Royal Greenwich Observatory (RGO) which is further extended by the NOAA. This combined RGO/NOAA data extends from 1874 May-2024 June.

We primarily use data recorded by the HMI onboard the Solar Dynamics Observatory (SDO), wherein we use the Space-weather HMI Active Region Patch (SHARP) data for calculating various physical properties of ARs \citep{Scherrer_2012_SolPhys, Pesnell_2012_SolPhys, Bobra_2014_SolPhys}. We analyze 33 magnetic parameters from the SHARP data, wherein 29 parameters are readily available as SHARP keywords and the remaining 4 parameters are manually calculated from SHARP images relying on vector magnetic field information (refer to Appendix \ref{sec:AppendixA}). For our analysis we specifically use hmi.sharp\_cea\_720s data series comprising of SHARP images starting from HMI AR patch (HARP) number 1 up to 11329, wherein all HARPs are associated with atleast one NOAA AR -- which sums to 1969 HARPs during SDO era over the period 2010 May-2024 June. For analyzing the total unsigned line-of-sight (LOS) magnetic flux (USFLUXL) of ARs we extend HMI database by including the observations from Michelson Doppler Imager (MDI) onboard the Solar and Heliospheric Observatory (SOHO) \citep{Scherrer_1995_Springer}, wherein we specifically use the Space-weather MDI Active Region Patch (SMARP) data \citep{Bobra_2021_ApJS}. The combined MDI/HMI database for total flux spans the solar cycles 23, 24 and ongoing cycle 25 till present over the period 1996 April-2024 June. For calibration details of total flux measurements refer to Appendix \ref{sec:AppendixB}.

\subsection{Does the size and flux of AR 13664 qualify it to be a super active region?}

\begin{figure*}
    \centering
    \includegraphics[width=1\linewidth]{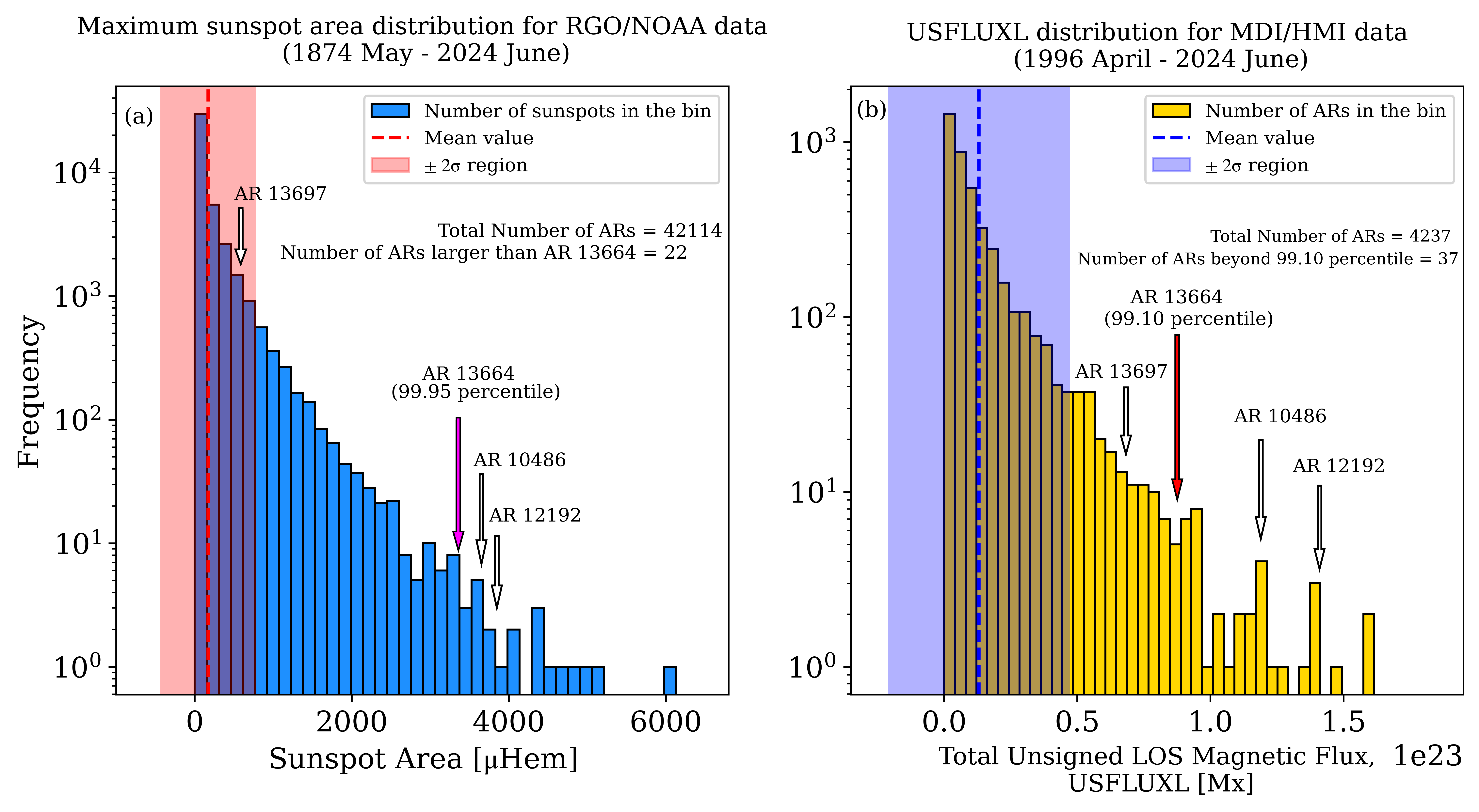}
    \caption{(a) Log scale histogram plot of maximum area of all the sunspot groups as recorded in RGO/NOAA database (1874 May-2024 June), wherein the area is measured in units of millionth of solar hemisphere (\(\mathrm{\mu}\)Hem). Red dashed line indicates the mean of all the maximum area values in the distribution, whereas red shaded region spans the mean \(\pm\) 2 \(\times\) standard deviation region. Pink arrow marks the position of RGO equivalent area of AR 13664 (3360 \(\mathrm{\mu}\)Hem) which stands at 99.95 percentile mark (only 22 out of 42114 ARs exceed AR 13664 in terms of size) in the distribution. (b) Log scale histogram of total unsigned line-of-sight (LOS) magnetic flux (USFLUXL) of ARs for the combined MDI and HMI database (1996 April-2024 June). We compute the flux content of ARs at the instance when their area is largest during their transit through the solar disk. Blue dashed line indicates the mean of total flux values and the blue shaded region spans the \(\pm\) 2 \(\times\) standard deviation region around the mean value. Red arrow indicates the position of total flux of AR 13664 (\(8.76\times10^{22}\) Mx) in the whole distribution which is stands at 99.10 percentile, depicting that only 37 out of 4237 ARs exceed the unsigned LOS flux content of AR 13664 in the combined MDI/HMI database. In both of the above distributions, two of the hollow arrows represent the position of area wise largest active regions of solar cycle 23 (AR 10486) and solar cycle 24 (AR 12192), whereas the third hollow arrow depicts the position of AR 13697 for visual reference. Note that 13697 is the NOAA identifier for AR 13664 during its second nearside transit.}
    \label{fig:Figure 3}
\end{figure*}

The maximum area of active pixels (AREA\_ACR, henceforth area) of AR 13664 during its lifetime is 4394.67 \(\mathrm{\mu}\)Hem (where \(\mathrm{1\ \mu}\)Hem \(\sim 3043675.5\ \mathrm{km^{2}}\), it would take around 104 Earths to cover the whole AR 13664) as measured by the SDO/HMI (refer to Figure \ref{fig:Figure 2}). As per the RGO/NOAA database, RGO equivalent area of AR 13664 is equal to 3360 \(\mathrm{\mu}\)Hem. Comparing this value of area with the distribution of maximum areas of all the sunspot groups in the RGO/NOAA database we find that AR 13664 ranks at 99.95 percentile, i.e., the top \(0.05\%\) over the period 1874 May-2024 June (refer to Figure \ref{fig:Figure 3}). From SHARP data we find the total unsigned LOS flux of AR 13664 to be equal to \(8.76 \times 10^{22}\) Mx. For the combined MDI and HMI era this value of flux stands at 99.10 percentile over the period 1996 April-2024 June. Note that during the second (Earth facing) transit of AR 13664 -- with re-designated NOAA number AR 13697 -- the area and flux had significantly reduced, emphasizing the role of AR decay and evolution in governing the activity of ARs. For comparison, relative to the largest ARs of the previous two solar cycles 23 and 24 (AR 10486 and AR 12192 respectively), AR 13664 ranks slightly lower in terms of size (refer to Figure \ref{fig:Figure 3}a). And the unsigned line-of-sight flux values for these two ARs (AR 10486 and AR 12192) exceed that of AR 13664's flux (refer to Figure \ref{fig:Figure 3}b). Our analysis show that only 22 out of 42114 ARs are larger than AR 13664 in terms of size over the period 1874 May-2024 June and only 37 out of 4237 ARs exceed its LOS magnetic flux content over the period 1996 April-2024 June. Given its large area, based on our analysis AR 13664 qualifies as a super active region \citep[SAR; see also][]{Chen_2011_aanda}.

\subsection{Properties and evolution of flare-relevant parameters of super AR 13664}
\begin{figure*}
    \centering
    \includegraphics[width=1\linewidth]{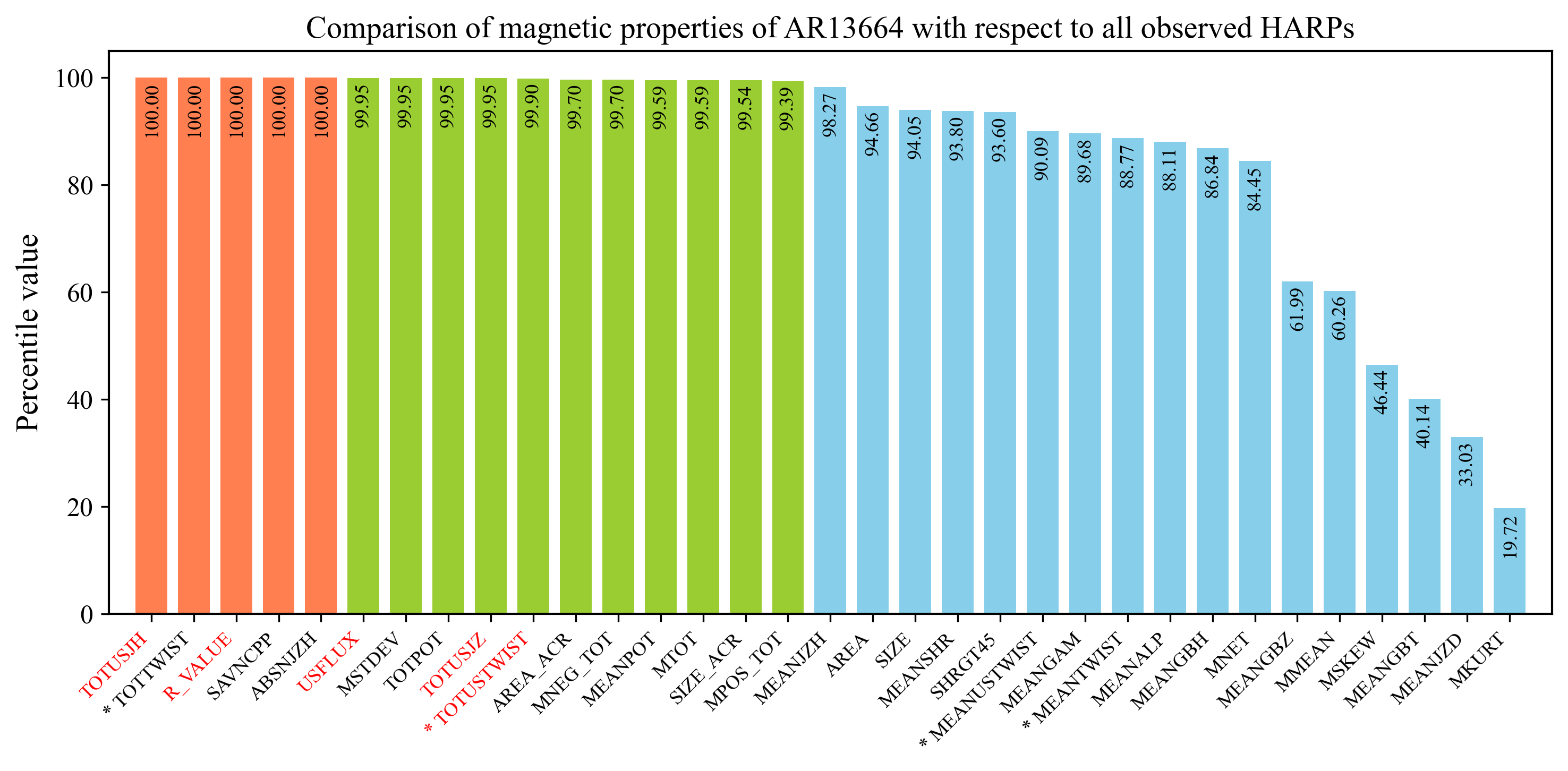}
    \caption{This bar plot compares percentile values of magnetic properties of AR 13664 at their peak with all observed NOAA associated ARs in the SHARP data series (HARP number 1-11329) over the period 2010 May-2024 June. The properties for all other ARs is considered at the time when their area is maximum during their respective disk transit. The magnetic parameters are horizontally arranged in order of decreasing percentile values from left to right. Red bars indicate percentile value of 100.00, whereas green bars represent percentiles between 99.00 and 99.99, followed by blue bars depicting percentiles less than 99.00. Red colored axis labels indicate that the parameter is one of the top five flare indicators as per the study by \cite{Sinha_2022_ApJ}. An asterisk before the parameter label symbolizes that the magnetic parameter is not a predefined SHARP keyword but rather it has been calculated manually (refer to Appendix \ref{sec:AppendixA}).}
    \label{fig:Figure 4}
\end{figure*}

We calculate various attributes of all ARs present in SDO/HMI SHARP database over the period 2010 May-2024 June to investigate how unique AR 13664 is, in general. While multiple observations are available for each AR, for this analysis we chose the SHARP data corresponding to the maximum projected area of each AR and we restrict ourselves to analyze only NOAA associated SHARP data. We find that five physical attributes of AR 13664 stand at 100 percentile mark in their respective distributions, implying that peak values of these specific properties for AR 13664 were largest among 1968 NOAA associated AR patches compared to the instances when their projected area peaked during their respective transits through the visible solar disk in the SDO era since 2010 (refer to Figure \ref{fig:Figure 4}). These parameters are total unsigned current helicity (TOTUSJH), net magnetic twist (TOTTWIST), total unsigned flux near the polarity inversion line (R\_VALUE), sum of absolute value of the net currents per polarity (SAVNCPP), and absolute value of the net current helicity (ABSNJZH). Out of the remaining ones eleven parameters lie in between 99.00-99.99 percentile range and the rest of them are below the 99.00 percentile mark.

\begin{figure*}
    \centering
    \includegraphics[width=1\linewidth]{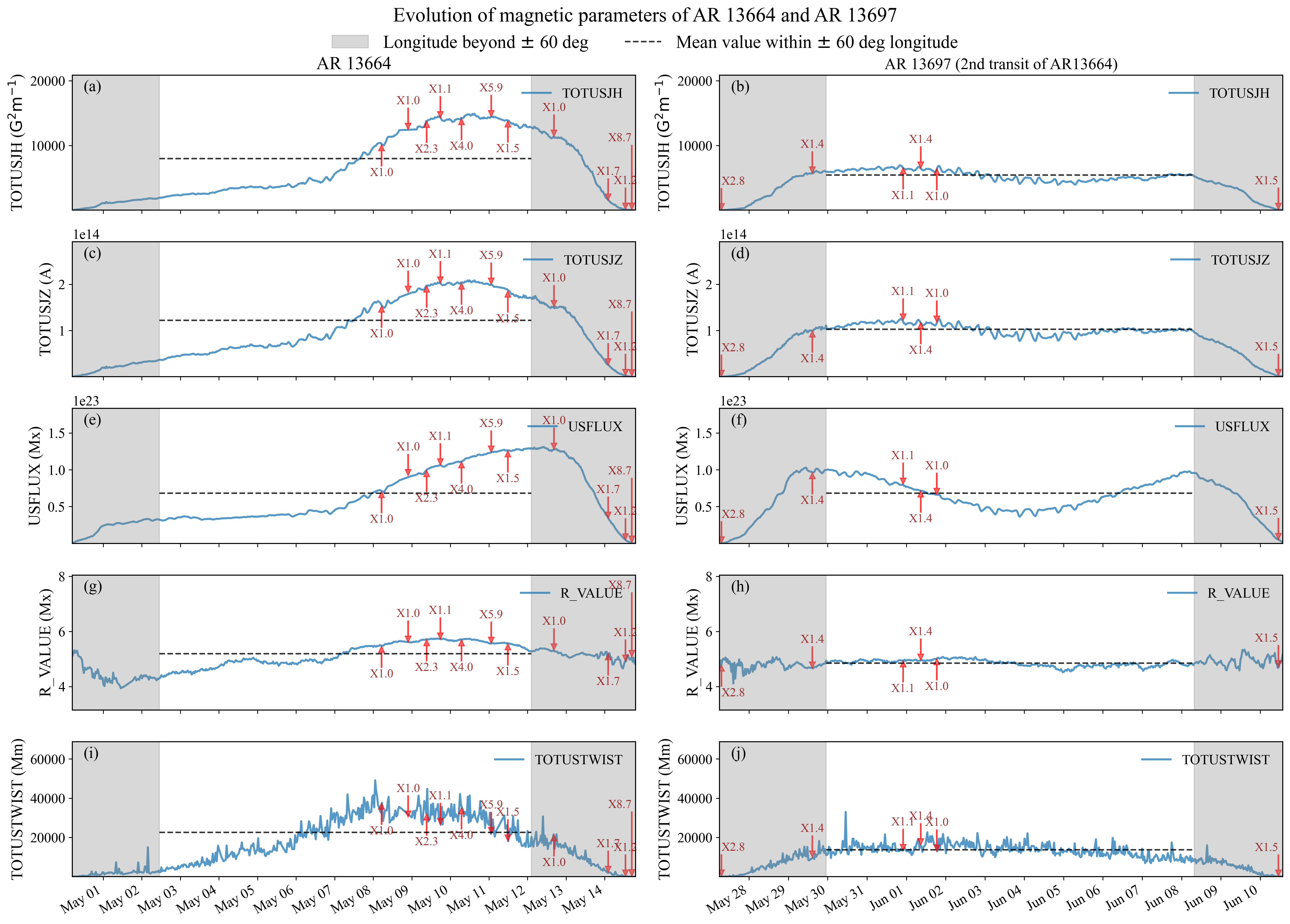}
    \caption{Temporal evolution (in blue) of (a)-(b) total unsigned current helicity (TOTUSJH), (c)-(d) total unsigned vertical current (TOTUSJZ), (e)-(f) total unsigned flux (USFLUX), (g)-(h) total unsigned flux near the polarity inversion line (R\_VALUE), and (i)-(j) total unsigned twist (TOTUSTWIST) for AR 13664 during its [Left] first nearside transit and [Right] second nearside transit (redesignated as AR 13697) through the solar disk. Gray shaded region spans the longitudes beyond \(-60^{\circ}\) and \(+60^{\circ}\) on the left and right end of the plots respectively. The mean value shown with a black dashed line indicates the time averaged value of AR's magnetic parameter during its transit between the chosen region of confidence of the solar disk (\(\pm 60^{\circ}\) longitude). Red arrows correspond to the time instances when AR 13664 produced an X-class flare, wherein the longest arrow points to the time of occurrence of the strongest flare produced by AR 13664, which is very close to the western limb of the Sun during the first nearside transit.}
    \label{fig:Figure 5}
\end{figure*}

Amongst various magnetic properties of an AR gleaned from HMI vector magnetogram SHARP data, \cite{Sinha_2022_ApJ} found that total unsigned current helicity (TOTUSJH), total unsigned vertical current (TOTUSJZ), total unsigned flux (USFLUX), total unsigned flux near the polarity inversion line (R\_VALUE), and total unsigned twist (TOTUSTWIST) are the top parameters contributing to the flare productivity \citep{Hahn_2005_ApJ}. These parameters are specifically identified in Figure \ref{fig:Figure 4} where we find that two out of the five of these flare relevant parameters rank at 100.00 percentile for AR 13664 in the SDO era. Notably, the top five fare relevant parameters as shown by \cite{Sinha_2022_ApJ} rank more than or equal to 99.90 percentile for this super AR 13664. Figure \ref{fig:Figure 5} depicts the temporal evolution of these five parameters and the area for the AR 13664 during its first and second nearside transits through the solar disk wherein the arrows mark the time of occurrence of X-class flares. We focus only on the region within \(\pm 60^{\circ}\) central meridian distance (CMD) because projection effects are likely to impact vector magnetic field observations near the limbs.

During the first nearside transit, we find that the flare relevant magnetic parameters under consideration increase to reach a peak -- around which most of the dynamic (X-class) flaring activity takes place (refer to Figure \ref{fig:Figure 5}). Notably, while the area of AR 13664 rises relatively smoothly over many days, we see evidence of a high rate of rise (i.e., rapid increase) in the total unsigned flux and other magnetic properties just before extreme flaring activity commences. This reinforces the hypothesis that adequate energization of the active region magnetic flux system is necessary for intense flaring activity to occur.

The frequency of flares tend to reduce between 2024 May 11-12, when most of the flare relevant parameters have started declining. However, we add the cautionary note that at this time AR 13664 is also about to move out of our region of confidence (within \(\pm 60^{\circ}\) CMD) and therefore, the estimate of parameter values beyond 2024 May 12 could be compromised. We note that the strongest X8.7 class flare produced by AR 13664 while it was facing the Earth, marked by the longest arrow in Figure \ref{fig:Figure 5}, is on 2024 May 14 when it is very near to the limb. It continued to produce more X-class flares in the far side of Sun which happened to be in the field of view of the Spectrometer/Telescope for Imaging X-rays (STIX) instrument onboard the European Space Agency's (ESA) Solar Orbiter (SolO) \citep{Krucker_2020_aanda, Muller_2020_aanda}. We combine the information from the STIX data center \citep{Xiao_2023_aanda} and ESA's recent science highlights\footnote{\url{https://www.esa.int/Science_Exploration/Space_Science/Solar_Orbiter/Can_t_stop_won_t_stop_Solar_Orbiter_shows_the_Sun_raging_on}} on this SAR to list all the X-class flares that emanated from AR 13664 on the solar farside. Out of these flares which occurred on the farside, one of the X-class flares recorded by the SolO has been estimated to have an intensity of X12 -- the stongest recorded solar flare of the current solar cycle. On 2024 May 15 AR 13664 rotated behind the western limb of the Sun and reappeared on 2024 May 27 to face the Earth again -- where it was re-designated as AR 13697. During its second nearside transit, AR 13664 produced 6 X-class flares in total, i.e., lower levels of flaring activity as compared to its first nearside transit. All the flare relevant parameters for AR 13664 peak in the first transit during its lifetime. Apart from the total unsigned flux, the time averaged mean values of the other four flare relevant parameters decrease during the second transit. This indicates a reduction in the energization of the original flux system associated with AR 13664. In total, AR 13664 produced 23 X-class flares over the period 2024 May 01-June 11 which is represented as a timeline of X-class flares during its first nearside transit, farside transit, and second nearside transit in Figure \ref{fig:Figure 1}.

As a comparative exercise we analyze the temporal evolution of top five flare productive parameters of AR 12192 -- the largest AR in terms of area in solar cycle 24 (refer to Appendix \ref{sec:AppendixA}). The AR 12192 produced six X-class flares during its Earth-facing transit. We surmise that this comparatively lower number of X-class flares could be because AR 12192 does not exhibit a similar rapid rise rate in the flare relevant parameters as AR 13664.

\section{Conclusions}

The impact of solar-stellar activity on planetary environments is a topic of great interest within the Sun-Earth system as well as exoplanetary systems. In particular, extreme events such as flares and coronal mass ejections have a profound effect on planetary atmospheres. In May this year, a gigantic magnetic active region on the Sun (AR 13664) unleashed a large number of highly energetic X class flares and associated CMEs; the properties of the latter are often influenced by attributes of their source regions and associated dynamical phenomena in the AR corona \citep{Pal_2017_ApJ}. The resulting Earth impact (geomagnetic storm) was the strongest in the last two decades. We perform the first comprehensive analysis of the magnetic properties of the active region that spawned these flares and identify this to be a super active region with very rare physical characteristics. We also demonstrate that intense flaring commences following a period of rapid energization of the system and persists across the phase where the flare productive physical properties remain high. Our findings for AR 13664 are consistent with the understanding that flux emergence and energization of AR associated overlying flux systems occur rapidly over a course of days, followed by slower decay and dispersal of the AR driven by turbulence and near surface flows. We conclude that the typical evolution of fundamental flare-associated physical parameters holds across a diverse range of flare productive ARs. An independent study focusing on free energy content of the same AR complex also indicates the most intense flaring activity begins after sufficient free energy build up \citep{Jarolim_2024_arXiv}.

We put AR 13664 in historical context by comparing it with all ARs which feature in RGO/NOAA and MDI/HMI observations spanning a cumulative period from 1874 May-2024 June. As noted earlier, this complex of activity involved merging of flux from two distinct NOAA AR identifiers (AR 13664 and AR 13668) before the flaring activity. They are so closely fused that it is difficult to delineate them and we surmise that they have erupted from the same underlying flux system.

We find that AR 13664 is characterized by extremely large values of area and total unsigned line-of-sight flux. Only 22 out of 42114 ARs in RGO/NOAA database exceed AR 13664 in terms of size and only 37 out of 4237 ARs exceed AR 13664 in the MDI/HMI database in terms of flux; i.e., AR 13664 is at the 99.95 percentile and 99.10 percentile mark in these categories respectively. Given its large area in terms of  both HMI measurements and RGO equivalence, AR 13664 meets and far exceeds the area criterion of being a super active region (SAR) as per the study by \cite{Chen_2011_aanda}.

Various physical attributes other than area and flux characterize an AR, some of these parameters also govern its flaring potential. Our analysis indicates that AR 13664 had multiple physical properties which rank at the top in the whole SDO era (since 2010 May-2024 June). Machine learning algorithms can be deployed to identify flare relevant parameters of an AR. Based on one such study which characterize flare relevant parameters of an AR \citep{Sinha_2022_ApJ}, our analysis shows that two out of the top five flare relevant parameters were the highest for AR 13664 and all five parameters were above the 99.00 percentile mark in the SDO era.

Finally, analyses of the temporal evolution of top ranking flare relevant physical attributes of AR 13664 show that extreme (X-class) flare activity is clustered around the peak values of these physical parameters -- reiterating that extreme activity levels require extreme energization of an AR flux system.

Taken together, our analysis establishes the extreme nature of AR 13664, firmly putting it at the high end of solar super active regions and demonstrates its uniqueness in terms of flare relevant parameters. We are near solar maximum conditions and expect to see a few more SARs over the course of the declining phase of sunspot cycle 25. Given the dynamic character and space weather relevance of such SARs, our study provides a paradigm to characterize them and explore physical attributes that lead to their extreme flare and CME productivity \citep[see also][]{Nandy_2003_ApJ, Pal_2018_ApJ, Sinha_2019_ApJ}.

The evolution of the dynamo mechanism of solar like star \citep{Nandy_2004_SolPhys, Hazra_2023_SpaceSciRev} due to changes in the solar interior and angular momentum losses mediated via solar winds point out that the Sun would have been a much more active star in its past \citep{Bindesh_2021_MNRAS}. Solar evolution over its lifetime would have manifested in both the nature and flux content of ARs and other solar activity levels \citep{Nandy_2007_AdvSpaceResearch, Nandy_2021_PEPS}. Thus, we may surmise that the Sun in its past history may have produced many more SARs, much more frequently than at present. Independent observations show evidence of super flares in other stars \citep{Maehara_2012_Nature, Notsu_2013_ApJ, Karoff_2016_NatComm} and also indicate that super flares may have occurred in the past history of the Sun \citep{Miyake_2012_Nature, Thomas_2013_GRL, Sakurai_2020_Nature}. It would be intriguing to explore how SARs bridge diverse scales of magnetically driven phenomena associated with an average AR and a super flaring AR. Our work illuminates how flare productive super active regions on the Sun and other stars may be identified and what are their salient physical properties.

\section*{acknowledgements}
CESSI is supported by IISER Kolkata, Ministry of Education, Government of India. P.J. acknowledges PhD fellowship from the Human Resource Development Group (HRDG) of Council of Scientific and Industrial Research (CSIR), Government of India under the file number 09/0921(19030)/2024-EMR-I. We thank Yoshita Baruah, Shaonwita Pal and an anonymous referee for useful discussions and comments that improved the manuscript.

\section*{Data Availability}

For our analysis we utilize the Royal Greenwich Observatory/USAF-NOAA active region database compiled by David H Hathaway (RGO/USAF/NOAA Data Centre 1874-2024\footnote{\href{https://solarscience.msfc.nasa.gov/index.html}{https://solarscience.msfc.nasa.gov/index.html}}). All the magnetic field data are either collected or calculated from SDO/HMI SHARP and SOHO/MDI SMARP data, maintained by the Joint Science Operation Center (JSOC\footnote{\href{http://jsoc.stanford.edu}{http://jsoc.stanford.edu}}). We acknowledge the Community Coordinated Modeling Center (CCMC) at Goddard Space Flight Center for the use of the DONKI catalog, \href{https://kauai.ccmc.gsfc.nasa.gov/DONKI/}{https://kauai.ccmc.gsfc.nasa.gov/DONKI/}. We acknowledge the use of data from GOES observatory maintained by NOAA. This research used version 5.0.0 (doi: \href{https://doi.org/10.5281/zenodo.8037332}{10.5281/zenodo.8037332}) of the SunPy open source software package \citep{Barnes_2020_ApJ}. This research used version 0.7.1 (doi: \href{https://doi.org/10.5281/zenodo.10440411}{10.5281/zenodo.10440411}) of the drms open source software package \citep{Glogowski_2019_JOSS}.

\facility{RGO/USAF-NOAA, SDO/HMI, SOHO/MDI, GOES}

\appendix
\section{Description of magnetic parameters}
\label{sec:AppendixA}

\begin{table*}[h]
    \centering
    \caption{Magnetic properties of AR along with their description has been tabulated below \citep{Bobra_2014_SolPhys}. The parameters are arranged from top to bottom in the decreasing order of their percentile values which represent their rank in their respective distribution for all the ARs in the SDO/HMI database over the period 2010 May-2024 June. AR 13664 survived long enough to face Earth the second time, where it was reidentified as AR 13697 by NOAA. It is to be noted that physical properties of ARs 13664 and 13697 are compared with the instances when the area of all other ARs is maximum during their respective disk transits. The top five flare relevant parameters as per the study by \cite{Sinha_2022_ApJ} are written in red color and an asterisk (*) symbol before the parameter denotes that the parameter has been manually calculated and is not one of the readily available SHARP keywords. For the mean and net properties, only magnitude is considered while comparing all the ARs.}
    \scalebox{.65}{
    \begin{tabular}{c c c c c c}
         \hline
         \hline
         S.No. & Magnetic parmeter & Description & Units & \multicolumn{2}{c}{Percentile value}\\
         \cline{5-6}
         & & & & transit 1  & transit 2\\
         & & & & [AR 13664] & [AR 13697]\\
         \hline
         1 & \textcolor{red}{TOTUSJH} & Total unsigned current helicity & G\(^2\)m\(^{-1}\) & 100.00 & 99.75\\
         2 & * TOTTWIST & Total twist & Mm & 100.00 & 99.70\\
         3 & \textcolor{red}{R\_VALUE} & Total unsigned flux near the polarity inversion lines & Mx & 100.00 & 99.64\\
         4 & SAVNCPP & Sum of absolute value of the net currents per polarity & A & 100.00 & 99.59\\
         5 & ABSNJZH & Absolute value of the net current helicity & G\(^2\)m\(^{-1}\) & 100.00 & 99.49\\
         6 & \textcolor{red}{USFLUX} & Total unsigned flux & Mx & 99.95 & 99.95\\
         7 & MSTDEV & Standard deviation of LOS flux density & G &  99.95 & 83.69\\
         8 & TOTPOT & Total photospheric magnetic energy denstiy & erg cm\(^{-3}\) & 99.95 & 98.27\\
         9 & \textcolor{red}{TOTUSJZ} & Total unsigned vertical current & A & 99.95 & 99.64\\
         10 & * \textcolor{red}{TOTUSTWIST} & Total unsigned twist & Mm & 99.90 & 99.70\\
         11 & AREA\_ACR & De-projected area of active pixels on sphere in micro-hemishpere & \(\mathrm{\mu}\)Hem & 99.70 & 99.54\\
         12 & MNEG\_TOT & Absolute value of total negative LOS flux & Wb &  99.70 & 99.39\\
         13 & MEANPOT & Mean photospheric excess magnetic energy density & erg cm\(^{-3}\) & 99.59 & 79.27\\
         14 & MTOT & Sum of absolute LOS flux within the identified region & Wb & 99.59 & 99.24\\
         15 & SIZE\_ACR & Projected area of active pixels on image in micro-hemisphere & \(\mathrm{\mu}\)Hem & 99.54 & 99.54\\
         16 & MPOS\_TOT & Absolute value of total positive LOS flux & Wb & 99.39 & 99.24\\
         17 & MEANJZH & Mean current helicity & G\(^2\)m\(^{-1}\) & 98.27 & 62.91\\
         18 & AREA & De-projected area of patch on sphere in micro-hemisphere & \(\mathrm{\mu}\)Hem & 94.66 & 98.32\\
         19 & SIZE & Projected area of patch on image in micro-hemisphere & \(\mathrm{\mu}\)Hem & 94.05 & 98.17\\
         20 & MEANSHR & Mean shear angle & \(\deg\) & 93.80 & 63.47\\
         21 & SHRGT45 & Area with shear angle greater than 45 as a percent of total area & \% & 93.60 & 61.74\\
         22 & * MEANUSTWIST & Mean unsigned twist & Mm\(^{-1}\) & 90.09 & 67.73\\
         23 & MEANGAM & Mean inclination angle, gamma & \(\deg\) & 89.68 & 56.15\\
         24 & * MEANTWIST & Mean twist & Mm\(^{-1}\) & 88.77 & 79.73\\
         25 & MEANALP & Mean twist parameter, alpha & Mm\(^{-1}\) & 88.11 & 57.98\\
         26 & MEANGBH & Mean value of the horizontal field gradient & G Mm\(^{-1}\) & 86.84 & 80.84\\
         27 & MNET & Net LOS flux within the identified region & Wb & 84.45 & 93.70\\
         28 & MEANGBZ & Mean value of the vertical field gradient & G Mm\(^{-1}\) & 61.99 & 68.70\\
         29 & MMEAN & Mean of LOS flux density & G & 60.26 & 68.70\\
         30 & MSKEW & Skewness of LOS flux denstiy & - & 46.44 & 38.67\\
         31 & MEANGBT & Mean value of the total field gradient & G Mm\(^{-1}\) & 40.14 & 66.82\\
         32 & MEANJZD & Mean vertical current density & mA m\(^{-2}\) & 33.03 & 72.97\\
         33 & MKURT & Kurtosis of LOS flux density & - & 19.72 & 19.51\\      
         \hline
    \end{tabular}}
    \label{tab:Table 1}
\end{table*}

\begin{figure*}
    \centering
    \includegraphics[width = 1\linewidth]{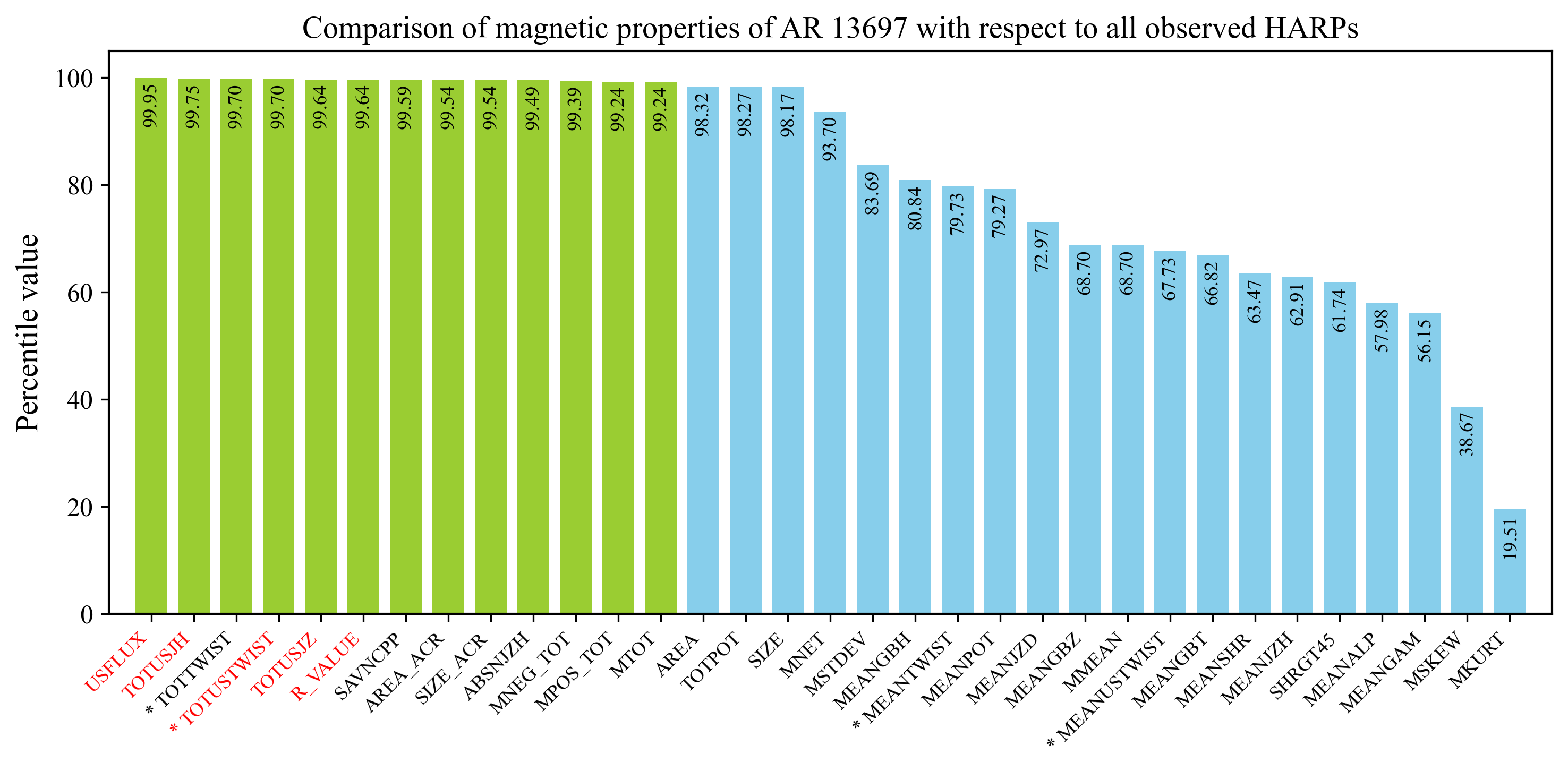}
    \caption{Comparison of percentile values of magnetic properties of AR 13664 at their peak with all observed NOAA associated ARs in the SHARP data series (HARP number 1-11329) over the period 2010 May-2024 June. The properties for all other ARs is considered at the time when their area is maximum during their respective disk transit. The magnetic parameters are horizontally arranged in order of decreasing percentile values from left to right. Green bars represent percentiles greater than or equal to 99.00, whereas blue bars depict the percentiles less than 99.00. Red colored axis labels indicate that the parameter is one of the top five flare indicators as per the study by \cite{Sinha_2022_ApJ}. An asterisk before the parameter label symbolizes that the magnetic parameter is not a predefined SHARP keyword but rather it has been calculated manually.}
    \label{fig:Figure A1}
\end{figure*}

\begin{figure*}
    \centering
    \includegraphics[width = 1\linewidth]{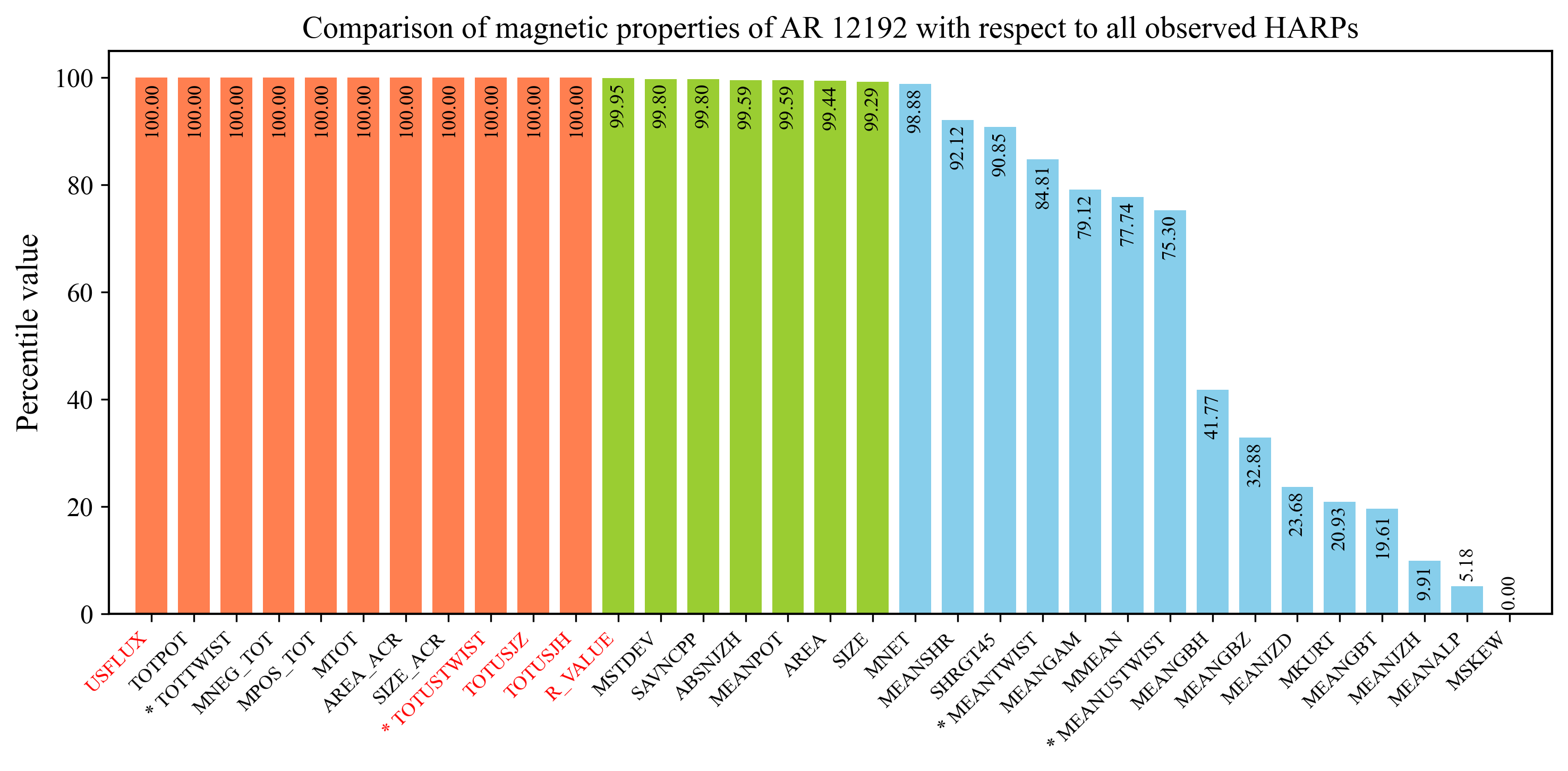}
    \caption{Comparison of percentile values of magnetic properties of AR 12192 at their peak with all observed NOAA associated ARs in the SHARP data series (HARP number 1-11329) over the period 2010 May-2024 June. The properties for all other ARs is considered at the time when their area is maximum during their respective disk transit. The magnetic parameters are horizontally arranged in order of decreasing percentile values from left to right. Red bars indicate percentile value of 100.00, whereas green bars represent percentiles between 99.00 and 99.99, followed by blue bars depicting percentiles less than 99.00. Red colored axis labels indicate that the parameter is one of the top five flare indicators as per the study by \cite{Sinha_2022_ApJ}. An asterisk before the parameter label symbolizes that the magnetic parameter is not a predefined SHARP keyword but rather it has been calculated manually.}
    \label{fig:Figure A2}
\end{figure*}

\begin{figure}
    \centering
    \includegraphics[width = 0.7\linewidth]{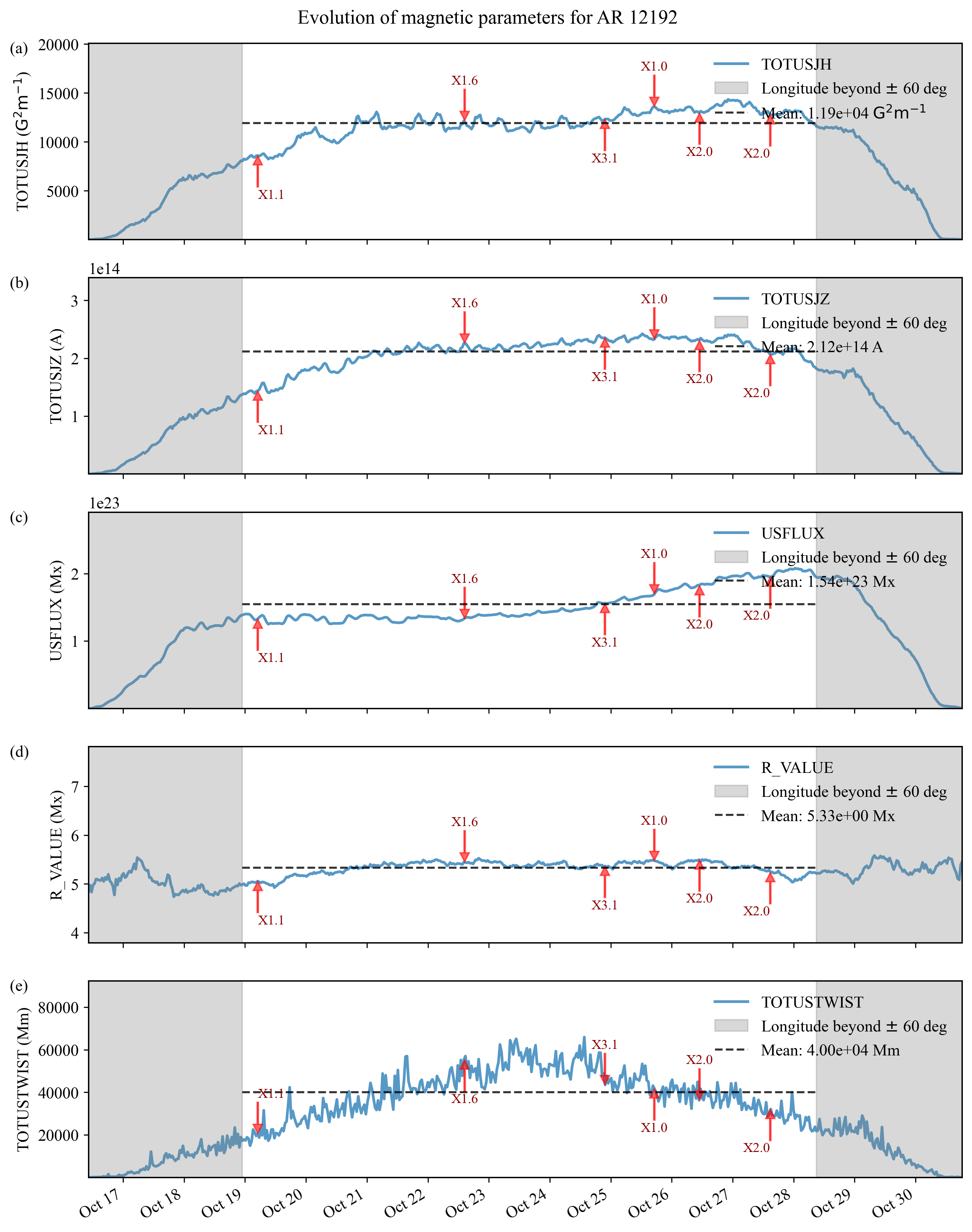}
    \caption{Temporal evolution (in blue) of (a) total unsigned current helicity (TOTUSJH), (b) total unsigned vertical current (TOTUSJZ), (c) total unsigned flux (USFLUX), (d) total unsigned flux near polarity inversion line (R\_VALUE), and (e) total unsigned twist (TOTUSTWIST) for AR 12192 during its transit through the solar disk. Gray shaded region spans the longitudes beyond \(-60^{\circ}\) and \(+60^{\circ}\) on the left and right end of the plots respectively. The mean value shown with a black dashed line indicates the time averaged value of AR's magnetic parameter during its transit between the chosen region of confidence of the solar disk (\(\pm 60^{\circ}\) longitude). Red arrows mark the time instances when AR 12192 produced an X-class flare}
    \label{fig:Figure A3}
\end{figure}

We analyze several physical properties of ARs using SHARP data to inspect how extreme AR 13664 is. Table \ref{tab:Table 1} contains all these properties along with their description, physical units and percentile values indicating their rank among all the ARs over the SDO era. It is to be noted that physical properties of ARs 13664 and 13697 are compared with the instances when the area of all other ARs is maximum during their respective disk transits. We reiterate the fact that AR 13664 survived long enough to face Earth the second time, where it was reidentified as AR 13697 by NOAA. To compute the temporal evolution of flare relevant parameters we use ``sharp\_cea\_720s" data series of HMI instrument \citep[refer to Figure \ref{fig:Figure 4};][]{Bobra_2014_SolPhys}. We use time averaged values of these parameters only within \(\pm 60^{\circ}\) CMD range to compute the mean value referred in Figure \ref{fig:Figure 5}. The formula for calculating time averaged mean \( \langle Q \rangle \) of a parameter \( Q \) is,
\[
\langle Q \rangle = \frac{1}{T} \sum_{i=1}^{N-1} Q_i (t_{i+1} - t_i)
\]
where $N$ is the number of data points in the time series and \( T = t_N - t_1 \) is the transit time of AR between \(-60^\circ\) and \(+60^\circ\) latitude. The reason behind choosing time averaged mean over the arithmetic mean is the presence of irregularity in the data-sampling due to occasional gaps in SHARP data. The physical properties which are not readily available as SHARP keywords and have been manually calculated using SHARP images are distinctly marked with an asterisk (*) in Table \ref{tab:Table 1}. The formulae for calculating these parameters -- TOTTWIST, MEANTWIST, TOTUSTWIST and MEANUSTWIST -- are as follows,

\begin{equation}
    \label{eq:A1}
    \mathrm{TOTTWIST} = \sum_{i=1}^{\mathrm{N_{pix}}} \left (\frac{\left(\nabla \times \mathrm{\textbf{B}}\right)_{\mathrm{z}}}{\mathrm{B_{z}}} \right ) \times \mathrm{dA}\\
\end{equation}

\begin{equation}
    \label{eq:A2}
    \mathrm{MEANTWIST} = \frac{1}{\mathrm{N_{pix}}} \times \sum_{i=1}^{\mathrm{N_{pix}}} \left (\frac{\left(\nabla \times \mathrm{\textbf{B}}\right)_{\mathrm{z}}}{\mathrm{B_{z}}} \right )\\
\end{equation}

\begin{equation}
    \label{eq:A3}
    \mathrm{TOTUSTWIST} = \sum_{i=1}^{\mathrm{N_{pix}}}\left|\frac{\left(\nabla \times \mathrm{\textbf{B}}\right)_{\mathrm{z}}}{\mathrm{B_{z}}}\right| \times \mathrm{dA}\\
\end{equation}

\begin{equation}
    \label{eq:A4}
    \mathrm{MEANUSTWIST} = \frac{1}{\mathrm{N_{pix}}} \times \sum_{i=1}^{\mathrm{N_{pix}}}\left|\frac{\left(\nabla \times \mathrm{\textbf{B}}\right)_{\mathrm{z}}}{\mathrm{B_{z}}}\right|
\end{equation}

where \(\mathrm{N_{pix}}\) and \(\mathrm{dA}\) are the total number of active pixels on which the parameters have been calculated and the physical area corresponding to one pixel in the CEA projected SHARP image respectively. For the mean and net parameters, only the magnitude of the parameters is considered while comparing all the ARs. Figure \ref{fig:Figure A1} represents the percentile rankings of different physical properties of AR 13697 (AR 13664 during second transit). There is visibly clear dip in the ranking of different physical properties such that none of the properties rank at 100.00 percentile. For comparison Figure \ref{fig:Figure A2} shows similar rankings for the largest active region of solar cycle 23 i.e., AR 12192. Additionally, similar to Figure \ref{fig:Figure 5}, we depict temporal evolution of top five flare relevant physical properties for AR 12192 in Figure \ref{fig:Figure A3}.

\section{Calibrating total unsigned line-of-sight flux for SOHO/MDI and SDO/HMI}
\label{sec:AppendixB}

\begin{figure}
    \centering
    \includegraphics[width=0.5\linewidth]{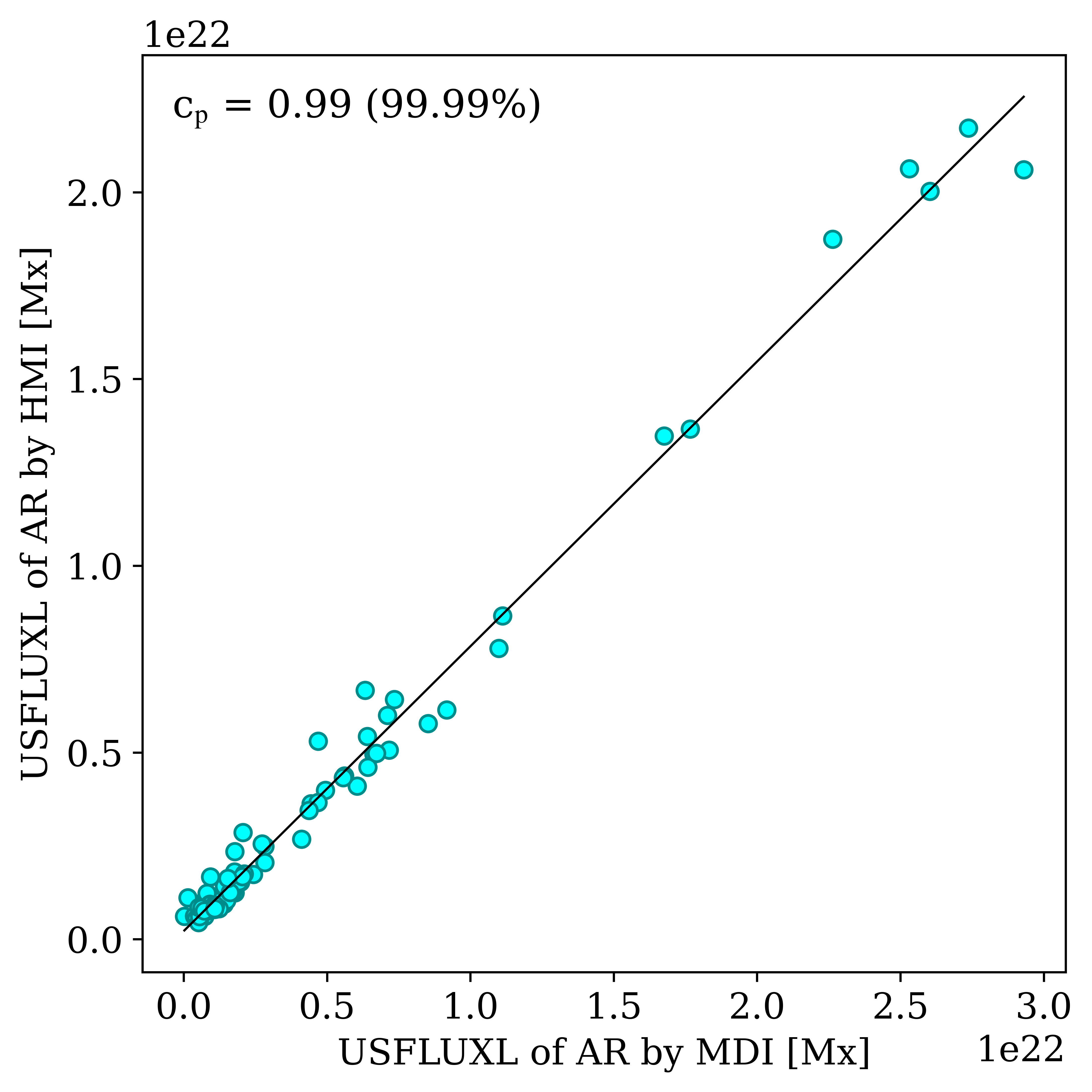}
    \caption{Scaling between the total unsigned line-of-sight (LOS) magnetic flux of active regions with identical NOAA number as recorded in the SDO/HMI and SOHO/MDI databases (marked as blue spots) for the period 2010 May-October. The black solid line represents the best fit curve with the relation \(\mathrm{USFLUXL_{HMI}[Mx]} = 0.762 \times \mathrm{USFLUXL_{MDI}[Mx]} + 2.165 \times 10^{20}\) where \(\mathrm{USFLUXL_{HMI}}\) and \(\mathrm{USFLUXL_{MDI}}\) are the values of total unsigned LOS flux of an AR as per SDO/HMI and SOHO/MDI respectively. \(c_{p} = 0.99\) is the pearson correlation coefficient between the two aforementioned physical quantities with 99.99\% confidence value.}
    \label{fig:Figure B1}
\end{figure}

Total unsigned LOS flux measurements of ARs by MDI and HMI overlap for the period 2010 May-October. During this time, the flux content of an AR in HMI and MDI data with identical NOAA number are associated by the following relation (refer to Figure \ref{fig:Figure B1}),

\begin{equation}
    \label{eq:B5}
    \mathrm{USFLUXL_{HMI}} = 0.762 \times \mathrm{USFLUXL_{MDI}} + 2.165 \times 10^{20}
\end{equation}

where \(\mathrm{USFLUXL_{HMI}}\) and \(\mathrm{USFLUXL_{MDI}}\) are the total unsigned LOS flux (in Mx) of an active region as measured by the SDO/HMI and SOHO/MDI respectively for the observation when the area of ARs is maximum during their transit on the solar disk. We use this relation to calibrate the flux values from the two aforementioned instruments and bring them at equal footing.

\newpage
\bibliography{bibliography_file}{}
\bibliographystyle{aasjournal}

\end{document}